\documentclass[useAMS,usenatbib]{mn2e}

\usepackage{graphicx}

\title[Two distant BDs in the UKIDSS DXS DR2]{
Two distant brown dwarfs in the UKIRT Infrared Deep Sky Survey Deep 
Extragalactic Survey Data Release 2
\thanks{Based on observations made with the United Kingdom Infrared
Telescope, the Canada-France-Hawaii Telescope Legacy Survey, and the 
Gemini Observatory}}
\author[N.\ Lodieu et al.]{N.\ Lodieu$^{1}$\thanks{E-mail: nlodieu@iac.es},
P.\ D.\ Dobbie$^{2}$, N.\ R.\ Deacon$^{3}$, B.\ P.\ Venemans$^{4}$, 
and M.\ Durant$^{1}$ \\
$^{1}$Instituto de Astrof\'isica de Canarias, V\'ia L\'actea s/n, 
E-38200 La Laguna, Tenerife, Spain \\
$^{2}$Anglo-Australian Observatory, P.O. Box 296, Epping 1710, 
Australia \\
$^{3}$Department of Astrophysics, Radboud University Nijmegen,
P.O. Box 9010, 6500 GL Nijmegen, The Netherlands \\
$^{4}$Institute of Astronomy, Madingley Road, Cambridge CB3 0HA, UK \\
}

\begin{document}

\date{Accepted \today. Received \today; in original form 23 June 2008}

\pagerange{\pageref{firstpage}--\pageref{lastpage}} \pubyear{2005}

\maketitle

\label{firstpage}

%
%
\begin{abstract}
We present the discovery of two brown dwarfs in the UKIRT Infrared Deep
Sky Survey (UKIDSS) Deep Extragalactic Survey (DXS) Data Release 2\@. 
Both objects were selected photometrically from six square degrees in DXS
for their blue $J-K$ colour and the lack of optical counterparts in the 
Sloan Digital Sky Survey (SDSS) Stripe 82. Additional optical photometry
provided by the Canada-France-Hawaii Telescope Legacy Survey (CFHT-LS)
corroborated the possible substellarity of these candidates. Subsequent 
methane imaging of UDXS J221611.51$+$003308.1 
and UDXS J221903.10$+$002418.2, has confirmed them as T7$\pm$1 
and T6$\pm$1 dwarfs at photometric distances 
of 81 (52--118 pc) and 60 (44--87 pc; 2$\sigma$ confidence level). 
A similar search in the second data release of the Ultra Deep Survey over 
a smaller area (0.77 deg$^{2}$) and shallower depth didn't return any 
late-T dwarf candidate. The numbers of late-T dwarfs in our study 
are broadly in line with a declining mass function when considering
the current area and depth of the DXS and UDS\@.
These brown dwarfs are the first discovered in the VIMOS 4 
field and among the few T dwarfs found in pencil-beam surveys. They 
are valuable to investigate the scale height of T dwarfs.
\end{abstract}


%
%
\section{Introduction}
\label{dT:intro}

T dwarfs are brown dwarfs whose spectral energy distribution is
mainly shaped by methane and water at near-infrared wavelengths
\citep{burgasser06a}. Their effective temperatures (T$_{\rm eff}$) are 
below 1400\,K \citep{golimowski04a,vrba04} and the coolest known to date, 
ULAS J133553.45+113005.2 has an estimated T$_{\rm eff}$ of
below 600\,K \citep{burningham08a}. The first T dwarf in the field was
discovered orbiting an M star, Gl229A \citep{nakajima95,oppenheimer95} 
and the same year \citet{rebolo95} discovered the first brown dwarf 
in the Pleiades. Thirteen years later, over 140 T dwarfs are known, 
the full list is available in an archive dedicated to low-mass stars 
and brown dwarfs\footnote{http://dwarfarchives.org,
a compendium of M, L and T dwarfs maintained by C.\ Gelino,
D.\ Kirkpatrick, and A.\ Burgasser.}, including 8 T7, 7 T7.5, and 3 T8,
and 3 T8.5--T9, following the classification at infrared wavelengths
\citep[][see the archive for the full list of objects and 
references]{burgasser06a}. However, most of these T dwarfs are nearby
and only two field T dwarfs have been announced in deep pencil-beam surveys:
a T7 in the NTT Deep Survey by \citet{cuby99} and T3--T4 dwarf in the 
IfA Deep Survey \citep{liu02a} along with a young T dwarf in the
$\sigma$ Ori cluster \citep{zapatero08a}. Those deep surveys provide
deep imaging over a limited area but represent ideal grounds to 
uncover various (and possibly new) types of objects (including
brown dwarfs) at large distances and perform science impossible
with shallower surveys. This work aims at finding distant and cool 
brown dwarfs with T$_{\rm eff}$ below $\sim$ 1000\,K and initiating 
a study of their Galactic distribution and scale height to improve the
mass function.

The UKIRT Infrared Deep Sky Survey 
\citep[UKIDSS;][]{lawrence07}\footnote{www.ukidss.org} is a new 
near-infrared survey conducted with the wide-field camera
\citep[WFCAM;][]{casali07} on the 4-m UK InfraRed Telescope 
(UKIRT). The UKIDSS photometric system, described in 
\citet{hewett06}, is based on the Mauna Kea Observatory 
system \citep{tokunaga02}. The data are 
pipelined-processed in the Cambridge Astronomical 
Survey Unit (CASU; Irwin et al., in prep)
and archived in the WFCAM Science Archive \citep[WSA;][]{hambly08}.
The project consists of three shallow surveys (Large Area Survey; LAS), 
Galactic Clusters Survey, and the Galactic Plane Survey) and two 
extragalactic components: the Deep Extragalactic Survey (DXS) and the 
Ultra-Deep Survey (UDS). 
One of the main scientific goals of UKIDSS, and the LAS component in 
particular, is to find the coolest and nearest brown dwarfs, in
particular those bridging the gap between the coolest T dwarfs and
planets \citep[so-called Y dwarfs;][]{kirkpatrick99}.
A dozen new T dwarfs have already been reported from a search in the
190 square degrees released in the LAS Data Release 1
\citep{kendall07,lodieu07b,chiu08}, including one T7 and two T7.5 dwarfs, 
as well as one of the coolest T dwarfs ever found 
\citep[ULAS0034$-$0052 classified as T8.5;][]{warren07c}. New 
discoveries are reported in \citet{pinfield08} and \citet{burningham08a}.
Although dedicated mainly to the understanding a galaxy formation at 
high-redshift and other extragalactic projects, the DXS and UDS represent 
valuable hunting grounds to look for objects with unique infrared colours 
like late-T dwarfs.

In this paper we present the discovery of two faint and distant T 
dwarfs from a simple photometric selection in the UKIDSS DXS 
(Sect.\ \ref{dT:selection}). Additional optical photometry was
obtained from the Canada-France-Hawaii Telescope (CFHT) Legacy Survey 
and the Sloan Digital Sky Survey (SDSS) deep stacks (known as Stripe 82) 
to remove contaminants over a significant area surveyed by the DXS 
Second Data Release (DR2). The photometric follow-up carried out in 
the methane ON (CH$_{4}$l) and OFF (CH$_{4}$s) narrow-band filters 
for both candidates confirmed them as a cool brown dwarfs 
(Sect.\ \ref{dT:photometry}). The spectral classification of these 
new brown dwarfs based on near-infrared photometry is examined in 
Sect.\ \ref{dT:classification}.
Finally, we discuss the implications of our discovery with respect to 
the expected number of T dwarfs in the DXS (Sect.\ \ref{dT:nb_DXS}) 
and report on a similar search in the UDS (Sect.\ \ref{dT:UDS_DR2}).

%
%
%
\begin{figure}
  \includegraphics[width=\linewidth]{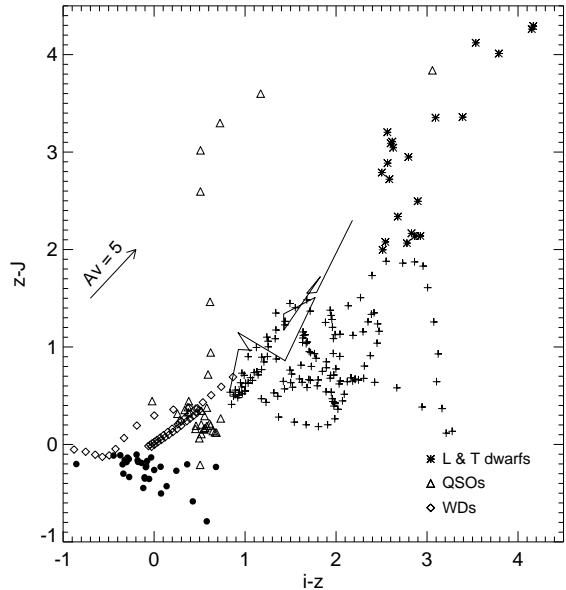}
   \caption{($i-z$,$z-J$) colour-colour diagram showing the location
of late-T candidates identified in the UKIDSS DXS DR2 and rejected
as not being brown dwarfs (filled circles). Overplotted are the synthetic
colours taken from \citet{hewett06} for M dwarfs (lines),
L and T dwarfs (star symbols), quasars (open triangles),
white dwarfs (open diamonds), and different types (Sa, Sb, Sc, E)
of redshifted Kinney--Mannucci galaxies (plus symbols).
}
   \label{fig_dT:2colour}
\end{figure}
%

%
%
\section{Sample selection}
\label{dT:selection}
\subsection{Selection in UKIDSS DXS DR2}
\label{dT:selection_DXSDR2}

The DXS goal is to cover 35 square degrees in four distinct fields
(Lockman Hole, XMM-LSS, VIMOS 4, and ELAIS N1) down to 5$\sigma$
depths of $J$ = 22.3 mag and $K$ = 20.8 mag for a point source
\citep{lawrence07}. 
In DR2, the approximate $J+K$ overlapping coverage is six square 
degrees and the achieved completeness limit is $K$ = 19.2-21.1 mag
because the survey is still on-going.
Observing conditions should meet the following criteria: seeing
better than 1.1 arcsec with thin cirrus or better \citep{dye06}.
The DXS fields have 10 second on-source integration times repeated 
several times to obtain intermediate stacks with a total exposure 
time of 640 seconds. These intermediate stacks are then repeated 
to achieve a final depth of $J$ = 22.3 mag and $K$ = 20.8 mag before 
moving to the next tile. 
The observed dispersion on the coordinates from the multiple epochs
can be used to estimate a proper motion. For both objects discussed in 
this paper, the time baseline span $\sim$6 months between June and
December 2005\@. The pipeline processing is identical
to the WFCAM standard reduction described in Irwin et al.\ (in prep).
No special treatment is applied to high proper motion sources
but a proper motion model should be implemented in UKIDSS
DR5 to look for moving sources (Nicholas Cross, personal communication).

We have input a simple Structure Query Language (SQL) query in WSA
to look for late-T dwarfs from the UKIDSS DXS Second Data Release 
\citep[DR2;][]{warren07b}. We have imposed the following constraints: 
point sources (ellipticity less than 0.333), stellar 
(i.e.\ the {\tt{mergedClass}} parameter between $-$2 and $-$1), 
upper limit of 0.5 arcsec on the separation/distance between the
two detections (parameters {\tt{jXi,kXi}} and {\tt{jEta,kEta}}),
and good quality ({\tt{ppErrBits}} parameter less than 
256)\footnote{see the WSA webpage at http://wsa.surveys.roe.uk for a 
detailed description of the parameters}. Note that the typical 
dispersion of the {\tt{jXi}}$-${\tt{kXi}} and {\tt{jEta}}$-${\tt{kEta}} 
parameters is low (because small offsets in the centroid between 
different bands are foreseen for the majority of sources),
corresponding to a 4--5 sigma clipping at the survey completeness limit.
In addition, we have limited our search
to $J$ fainter than 13 mag (to avoid saturated sources) and brighter 
than $K$ = 21.1 mag (5$\sigma$ limit of the DXS DR2). Finally, we have 
imposed a constraint on the infrared colour, namely 
$J-K \leq -$0.1 mag to select the bluest and therefore the coolest
T dwarfs. The query returned 54 sources. Nevertheless, additional
optical photometry is required to clean the sample of spurious 
candidates with neutral infrared colours. 

%
%
%
%
\begin{table*}
 \centering
 \caption[]{Coordinates, epoch of observations, and infrared photometry 
 of the two new T dwarfs extracted from the UKIDSS Deep Extragalactic Survey 
 Second Data Release. The first line for each object gives the DXS coordinates
 whereas the second line gives the CFHT coordinates.
 Optical photometry measured on the SDSS Stripe 82 (limits at 3$\sigma$)$^{a}$ and
 on the CFHT images$^{b}$ is given for both in the first and second line, respectively.
 The $JHK$ magnitudes are in the MKO system \citep{tokunaga02,hewett06}
 and the optical $i$ and $z$ magnitudes are in the AB system \citep{york00}}
 \begin{tabular}{l c c c c c c c c}
 \hline
UDXS J\ldots{} & RA   & dec   &  Epoch  & $J$   &  $K$  & $J-K$ & $i$ & $z$ \cr
\hline
221611.51$+$003308.1 & 22 16 11.51 & $+$00 33 08.1 & 2005-09-11 & 20.193$\pm$0.040 & 20.908$\pm$0.197 & $-$0.715 & $>$24.0        & 23.0$\pm$0.6$^{a}$ \cr
                     & 22 16 11.52 & $+$00 33 08.2 & 2006-09-17 &                  &                  &          & $>$26.1        & 24.42$\pm$0.29$^{b}$ \cr
221903.10$+$002418.2 & 22 19 03.10 & $+$00 24 18.2 & 2005-06-20 & 19.000$\pm$0.017 & 19.183$\pm$0.040 & $-$0.182 & $>$24.2        & $>$22.5$^{a}$ \cr
                     & 22 19 03.09 & $+$00 24 17.8 & 2006-08-11 &                  &                  &          & 26.43$\pm$0.61 & 23.42$\pm$0.10$^{b}$ \cr
\hline
 \label{tab_dT:list_cand}
 \end{tabular}
\end{table*}
\subsection{Optical photometry from public databases}
\label{dT:selection_opt}

First of all, we have input all 54 candidates from DXS DR2 into the 
SDSS DR6 database \citep{adelman_mccarthy08}\footnote{http://cas.sdss.org/dr6/en/tools/crossid/upload.asp}
and requested the nearest SDSS sources within 3 arcsec of the DXS
detection. A total of 36 sources were returned and none of them had
$z-J$ and $i-z$ colour typical of T dwarfs \citep{knapp04}. Thus, we have 
rejected those sources and are left with 54$-$36 = 18 candidates.
Fig.\ \ref{fig_dT:2colour} shows the location of these contaminants
in a ($i-z$,$z-J$) two-colour diagram, suggesting that they are likely
early-type stars according to the synthetic colours compiled
by \citet{hewett06}.

Second, we have looked into public databases with common coverage
to the DXS to remove potential outliers, including the CFHT Legacy Survey
(CFHTLS; see Sect.\ \ref{dT:selection_opt_CFHT})\footnote{http://www.cfht.hawaii.edu/Science/CFHTLS/}, the Subaru Deep 
Survey\footnote{http://www.naoj.org/Science/SubaruProject/SDS/},
and the SDSS Stripe 82 \citep{frieman08,adelman_mccarthy08}.
The SDSS stacks centered on the location of the UKIDSS objects were 
created by combining (on average) 23 single-epoch SDSS images.
The single-epoch SDSS images have been released as DRSN1 \citep{sako05}
and DRsup \citep{adelman_mccarthy07}. The SDSS image were scaled to a 
common zero-point using the SDSS DR6 catalogue and combined using standard 
routines in IRAF\@. 

Details of the location of the candidates and their photometry are 
as follows:
\begin{itemize}
\item For the three candidates located in XMM-LSS field centered at 
(RA,dec)=(02$^{\rm h}$25,$-$04$^{\circ}$30), we have extracted optical ($i+z$)
photometry from the CFHTLS Wide 1\@. All of them were clearly detected
and exhibit optical-to-infrared colours inconsistent with T dwarfs
\citep{knapp04}.
\item In the Lockman Hole pointing (10$^{\rm h}$57,$+$57$^{\circ}$40),
we have three candidates: one has an optical counterpart and no object
was visible on the $J$ and $K$ images of the other two candidates so 
we discarded them all.
\item In ELAIS N1 centered at (16$^{\rm h}$10,$+$54$^{\circ}$00), we 
have two candidates but both of them were artefacts beside a bright
star (known as cross-talks) and thus rejected.
\item Finally, the largest number of candidates (10 in total) is found 
in the VIMOS 4 pointings (22$^{\rm h}$17,$+$00$^{\circ}$20).
Part of that DXS VIMOS 4 field overlaps with the SDSS Stripe 82
in the $-$1.266$^{\circ}$ to $+$1.266$^{\circ}$ declination range. However 
the northern part of VIMOS 4 has currently no optical photometric catalogue
linked to it (from dec=1.266$^{\circ}$ to 1.8$^{\circ}$). Four candidates
lie in this area and three of them are detected in USNO \citep{monet03}
confirmed as contaminants by the CFHT photometry 
(Sect.\ \ref{dT:selection_opt_CFHT}).
The one without photometry, UDXS J222203.56$+$013330.2 (UDXS2222), is not 
included in Table \ref{tab_dT:list_cand} because discarded after inspection
of the CFHT 
photometry (Sect.\ \ref{dT:selection_opt_CFHT}). Three of the remaining 
six candidates are cross-talks and another one is clearly visible on the 
images of the SDSS Stripe 82 (thus discarded). The remaining two were kept 
as potential T dwarfs, UDXS J221611.51$+$003308.1 (hereafter UDXS2216) and 
UDXS221903.10$+$002418.2 (UDXS2219), because of their lower limits in 
$i$ and $z$ from the stacked images of the SDSS Stripe 82 
(Table \ref{tab_dT:list_cand}).
\end{itemize}
\subsection{Optical photometry from CFHT}
\label{dT:selection_opt_CFHT}

Photometrically calibrated stacks of CFHT/MegaCam images of the candidates 
were downloaded from the Canadian Astronomy Data 
Centre\footnote{http://www1.cadc-ccda.hia-iha.nrc-cnrc.gc.ca/cadc/}. 
Objects in a 6$\times$6 arcmin$^{2}$ box surrounding the object were 
matched with the UKIDSS catalogue using a search radius of 2 arcsec. 
Using the UKIDSS positions of the matched objects, the astrometry
of those CFHT images was updated. The resulting uncertainty between the 
positions of objects on the CFHT images and the UKIDSS catalogue is 
less than 0.1 arcsec. 

Magnitudes were measured in apertures with a diameter 1.5 times the seeing 
FWHM. Nearby bright stars were used to measure the aperture corrections. 
Magnitude uncertainties were computed by randomly placing apertures on the 
image and determining the dispersion of the flux distribution. UDXS2216 was 
imaged for 8610 sec in $i$ with a mean seeing of 0.67$"$ and for 3600 sec 
in $z$ with an average seeing of 0.71$"$. UDXS2219 was imaged for 4305 sec in 
$i$ and 10260 sec in $z$ with an image quality of 0.73$"$ and 0.80$"$, 
respectively. Finally UDXS2222 was only imaged in $z$ for 3600 sec with 
a mean seeing of 0.86$"$.

The transmission curves of the MegaCam $i$ and $z$ filters differ from
the SDSS filter curves: the MegaCam $i$ is redder than SDSS $i$ and the
MegaCam $z$ is bluer than the SDSS $z$, yielding colours with less
contrast. This comparison is detailed in \citet{delorme08b}. Their
Figures 3 and 7 show the $i-z$ colours as a function of spectral type
and $(z-J)_{\rm AB}$, respectively. Note that the $z-J$ colour is
in AB magnitudes (usually $z$ is in AB system and $J$ in the Vega
system). We will use this definition only for the three objects with CFHT 
photometry to assign tentative spectral types.

Firstly, UDXS2222 was clearly detected on the CFHT images and we measured
$z$ = 21.64$\pm$0.03 mag, implying $z-J$ of about 0.75, inconsistent with
T dwarfs \citep{delorme08b}. Therefore, we rejected this object as a
potential T dwarf. Secondly, we looked into the $i$ and $z$ magnitudes
for the remaining two T dwarf candidates with lower limits from the
SDSS Stripe 82\@. For UDXS2216, we measured a lower limit of 
$i >$ 26.1 and $z$ = 24.42$\pm$0.29 (Table \ref{tab_dT:list_cand}), 
implying $i-z >$ 1.6 and $(z-J)_{\rm AB}$ = 3.29 mag. Those values 
suggest a possible late-T 
dwarf according to Fig.\ 7 in \citet{delorme08b} and methane imaging 
presented in Sect.\ \ref{dT:photometry} confirms this hypothesis.
For UDXS2219, we measured $i$ = 26.4$\pm$0.6 and $z$ = 23.42$\pm$0.10
(Table \ref{tab_dT:list_cand}), yielding $(z-J)_{\rm AB}$ = 3.48 and 
$i-z \sim$ 3 mag placing that object in the region of late-T dwarfs 
in Fig.\ 7 of \citet{delorme08b}. Therefore, we consider both objects 
as late-T dwarf candidates. We note that UDXS2219 is redder than any
of the new T dwarf candidates presented in \citet{delorme08b}.

From the time difference of $\sim$1 year between the DXS and CFHT
observations, we have attempted to compute the proper motion for
both candidates. For UDXS2216 the motion is within the error bars
of the astrometry. The methane observations described in
Sect.\ \ref{dT:photometry} and taken two years after the DXS
images support a negligible proper motion since the measurements
are within the error bars. Repeating the same procedure for 
UDXS2219 suggests a proper motion 2.5$\sigma$ above the
astrometric errors (0.35$\pm$0.15 arcsec/yr).

%
%
%
\begin{table*}
 \centering
 \caption[]{Coordinates from the NIRI images and photometry of 
 the new late-T dwarfs, UDXS2216 (top) and UDXS2219 (bottom),
 and three bright stars close to the targets to measure the difference 
 between the photometry in the CH$_{4}$s and CH$_{4}$l filters. Note
 that the photometry is measured with an aperture of r=8 pixels, 
 equivalent to twice the
 measured seeing. The magnitude scale is arbitrary as we are only 
 interested in the relative colour. The final (CH$_{4}$s $-$ CH$_{4}$l) 
 colours measured for UDXS2216 ($-$1.24$\pm$0.15 mag) and UDXS2219 
 ($-$0.85$\pm$0.11 mag) translate into spectral types of T7$\pm$1 and 
 T6$\pm$1, respectively 
 \citep{tinney05}.}
 \begin{tabular}{c c c c c c c c c}
 \hline
Name  &  RA   & dec   &   $J$  &  $K$  &  CH$_{4}$s  & CH$_{4}$l  &  CH$_{4}$s$-$CH$_{4}$l  \cr
UDXS J\ldots{} & \multicolumn{2}{c}{J2000}     & \multicolumn{2}{c}{DXS photometry} & \multicolumn{2}{c}{aperture photometry} & \multicolumn{2}{c}{methane colour} \cr
\hline
221611.49$+$003307.8  & 22 16 11.49 & $+$00 33 07.8 & 20.193$\pm$0.040 & 20.908$\pm$0.197 & 17.008$\pm$0.053 & 18.472$\pm$0.142  & $-$1.464$\pm$0.152 \cr
221612.23$+$003258.8  & 22 16 12.23 & $+$00 32 58.8 & 17.968$\pm$0.010 & 17.142$\pm$0.010 & 14.331$\pm$0.005 & 14.573$\pm$0.005  & $-$0.242$\pm$0.005  \cr
221611.37$+$003317.8  & 22 16 11.37 & $+$00 33 17.8 & 18.389$\pm$0.012 & 17.507$\pm$0.012 & 14.741$\pm$0.007 & 14.963$\pm$0.006  & $-$0.222$\pm$0.007 \cr
221609.59$+$003322.5  & 22 16 09.59 & $+$00 33 22.5 & 18.206$\pm$0.011 & 17.410$\pm$0.014 & 14.557$\pm$0.006 & 14.781$\pm$0.006  & $-$0.224$\pm$0.006 \cr
\hline
\hline
221903.10$+$002418.2  & 22 19 03.10 & $+$00 24 18.2 & 19.000$\pm$0.017 & 19.183$\pm$0.040 & 21.290$\pm$0.040 & 21.954$\pm$0.044  & $-$0.664$\pm$0.059 \cr
221900.07$+$002423.3  & 22 19 00.07 & $+$00 24 23.3 & 19.400$\pm$0.021 & 18.583$\pm$0.025 & 21.310$\pm$0.040 & 19.874$\pm$0.014  & $+$0.226$\pm$0.042 \cr
221903.57$+$002443.7  & 22 19 03.57 & $+$00 24 43.7 & 17.905$\pm$0.009 & 17.545$\pm$0.012 & 19.958$\pm$0.016 & 19.874$\pm$0.014  & $+$0.084$\pm$0.021 \cr
221901.19$+$002416.8  & 22 19 01.19 & $+$00 24 16.8 & 18.605$\pm$0.013 & 17.724$\pm$0.013 & 20.563$\pm$0.024 & 20.299$\pm$0.017  & $+$0.264$\pm$0.029 \cr
\hline
 \label{tab_dT:phot}
 \end{tabular}
\end{table*}
%

%
%
%
\section{Methane photometric follow-up}
\label{dT:photometry}
\subsection{Observations and data reduction}

To confirm the substellar nature of our candidates,
we have obtained images in the CH$_{4}$l and CH$_{4}$s filters
available on the Gemini Near-Infrared Imager \citep[NIRI;][]{hodapp03}. 
This represents a very efficient method to infer a spectral type (for
T4 or later) in a reasonably short amount of telescope time and it is 
much faster in terms of observing time than obtaining a near-infrared 
spectrum for sources as faint as those extracted from the DXS\@.

NIRI was used in imaging mode with the f/6 camera giving a 0.117 arcsec
pixel scale and a field-of-view of 120 arcsec. Observations were made on 
in queue mode on 2007 August 08 (program GN-2007B-Q-88) and on 2008
September 02 (program GN-2008B-Q-90) for UDXS2216 and UDXS2219,
respectively. For the CH$_{4}$s filter (central wavelength is 1.58 $\mu$m) 
where the late-T dwarfs are the brightest (i.e., less affected
by the methane absorption band), we have used shorter 
integrations than for the CH$_{4}$l filter (centered at 1.69 $\mu$m). 

%
%
%
\begin{figure}
  \centering
  \includegraphics[width=\linewidth]{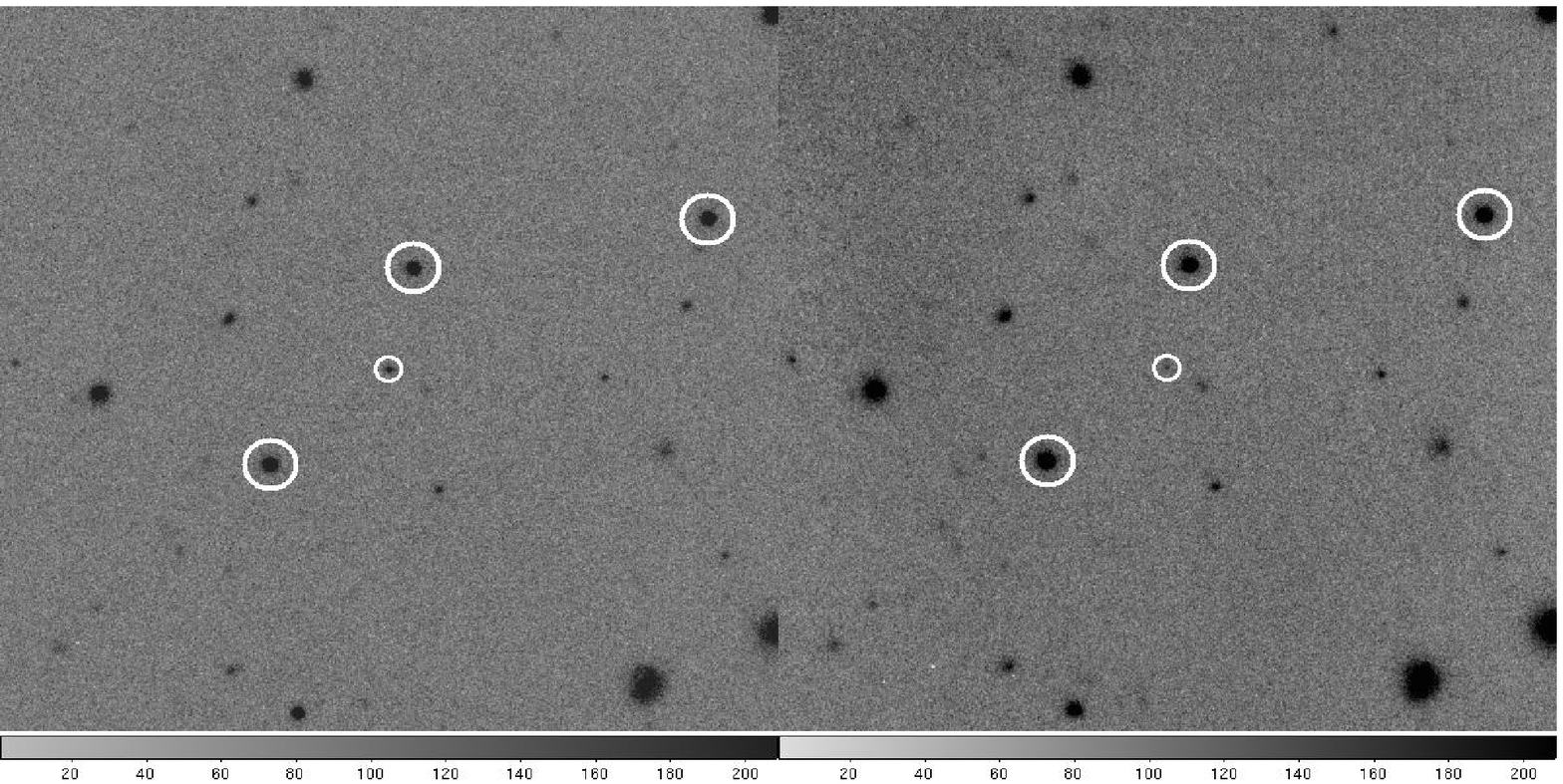}
  \includegraphics[width=\linewidth]{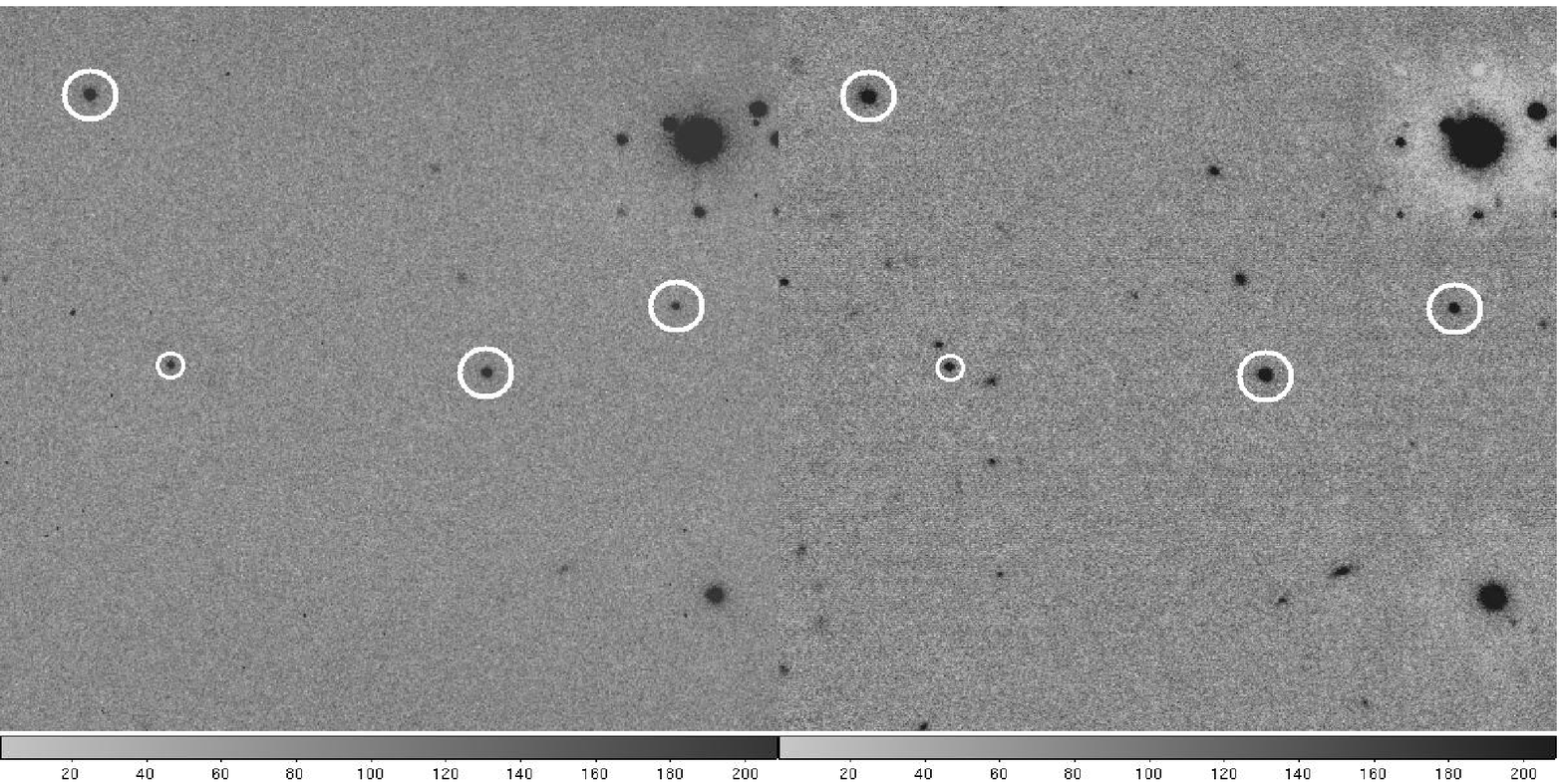}
   \caption{Images of UDXS2216 (top) and UDXS2219 (bottom) in the
CH$_{4}$s (left) and CH$_{4}$l (right) narrow-band filters. Targets
are circled with small circles. Each image is 70 arcsec aside
with North up and East to the left. Reference stars used to derive 
the methane index are marked with a large circle.
}
   \label{fig_dT:fc_target}
\end{figure}

For UDXS2216, we have used on-source integrations of 140 (62) sec 
repeated one (three times) and 10 (20) dithers following a standard 
pattern, yielding total exposure times of $\sim$23 min and 62 min in 
the CH$_{4}$s and CH$_{4}$l filters, respectively.
For UDXS2219, we have used on-source integrations of 55 (50) sec
and 9 (54) dithers, producing total exposure times of 8.25 min and 
45 min in the CH$_{4}$s and CH$_{4}$l filters, respectively.
The observing conditions for both objects were affected with some patchy 
clouds and seeing measured on the average images was below 0.5 arcsec
in both filters, respectively. The conditions were poorer for
UDXS2219 than for UDXS2216, resulting in larger uncertainty in the
photometric calibration. No photometric standard star was observed 
because we were primarily interested in relative photometry.

Data reduction was done using tasks in the Gemini IRAF NIRI
package following a standard procedure. Files were prepared and 
corrected for offset bias using NSPREPARE\@. Then, a normalised
flat-field was constructed using flats with the shutter on and off
as well as short darks to identify bad pixels. Afterwards, each science 
image was sky-subtracted and divided by the normalised flat-field.
A sky frame was created for each science frame after identifying
and removing the stars present on the image. Finally, all images
were co-added to create the final science frame. The procedure
was identical for both filters. The final images in the CH$_{4}$s 
and CH$_{4}$l filters are displayed in Fig.\ \ref{fig_dT:fc_target} 
(UDXS2216 top and UDXS2219 bottom) and the targets marked
with a small circle.

\subsection{Differential methane photometry}

We have carried out differential methane imaging (CH$_{4}$s $-$ CH$_{4}$l)
following the technique described in \citet{tinney05}. The photometry 
is neither corrected for exposure time and nor 
calibrated since we are only interested in relative photometry.
We have compared the colour of our targets using several bright 
nearby stars within the NIRI field-of-view (marked with large circles 
in Fig.\ \ref{fig_dT:fc_target}). We have assumed a zero methane colour 
for those stars, a reasonable assumption for any source with a spectral type
from A0 to T2 \citep{tinney05}.

From a visual inspection of the images, UDXS2216 appears much
fainter in the CH$_{4}$l filter, suggesting that it is indeed a
methane dwarf (Fig.\ \ref{fig_dT:fc_target}). To quantify the difference 
in magnitude and estimate its spectral type, we have measured the flux 
of UDXS2216 within an aperture radius r=8 pixels, corresponding to 
0.95 arcsec on the sky or twice the
measured full width at half maximum. We have repeated this procedure
for three bright stars within the NIRI field-of-view. The magnitudes
and their associated errors are listed in Table \ref{tab_dT:phot}. 
The mean offset in colour (or methane index) for the three reference
stars is $-$0.23 mag. 
Therefore, this value should be subtracted from the measured value for 
UDXS2216, yielding a final (CH$_{4}$s $-$ CH$_{4}$l) colour of $-$1.23 mag.
The errors on the final colour is $\sim$0.15 mag and is
dominated by the measurement uncertainties on the faint CH$_{4}$l magnitude 
i.e.\ where the methane absorption band is affecting the spectral energy
distribution of late-T dwarfs.

We have repeated the same procedure for UDXS2219 and three reference
stars in the NIRI field-of-view (Fig.\ \ref{fig_dT:fc_target}) using
an aperture of 8 pixels (as for UDXS2216 for consistency). We have
measured a mean offset of $+$0.191 mag (with a standard deviation of
0.09 mag) for the three stars and a methane
index of $-$0.664 mag for the target (Table \ref{tab_dT:phot}), implying 
a final colour of $-$0.85$\pm$0.11 mag. The photometric error include the 
measurement uncertainties on the CH$_{4}$s and CH$_{4}$l magnitudes as 
well as a calibration uncertainty from the reference stars.

%
%
%
\section{Spectral classification}
\label{dT:classification}

UDXS2216 is one of the bluest late-T dwarfs in $J-K$ 
\citep[values in the MKO system;][]{tokunaga02} 
extracted from UKIDSS: the bluest one published in \citet{lodieu07b} 
is ULAS J0222$+$0024, a T5 dwarf with $J-K$ = $-$0.47 mag. Since then, 
new mid to late-T dwarf have been discovered with bluer $J-K$ colours 
down to $-$0.95 mag \citep{pinfield08}.
We observe a significant dispersion among T dwarfs from UKIDSS LAS 
in the $J-K$ vs spectral type relation, suggesting that our new candidate
is later than T5\@. According to Figure 6 of \citet{pinfield08}
a $J-K$ colour of $-$0.71 mag would translate into a spectral type of 
T5--T5.5 or T7--T8.5\@.
To clarify this issue, we have investigated the synthetic colours
of T dwarfs published by \citet{hewett06}. Among the coolest T dwarfs
listed in their Table 10, there is a clear cut-off in the $J-K$ 
colours between T4.5 and T6 with all sources later than T6 being
bluer than $-$0.31 mag, the bluest being Gl570D 
\citep[T7.5;][]{burgasser00a,burgasser06a} with $J-K$ = $-$0.71 mag.
Hence, according to synthetic colours \citep{hewett06}, 
UDXS2216 is likely to be a T7.5 dwarf, in agreement with
the latest spectral range derived from observed colours of LAS T dwarfs.

The $J-K$ colour of UDXS2219 is not as blue as UDXS2216, placing
less constraints on the spectral type. Again, according to Figure 6
of \citet{pinfield08}, UDXS2219 could be an early-T dwarfs but also
a mid to late-T with spectral type between T5 and T6.5\@. As pointed
out earlier, the optical-to-infrared colour, however, places UDXS2219 
in the late-T regime \citep{delorme08b}, favouring the latest spectral 
types.

\begin{figure}
  \includegraphics[width=\linewidth]{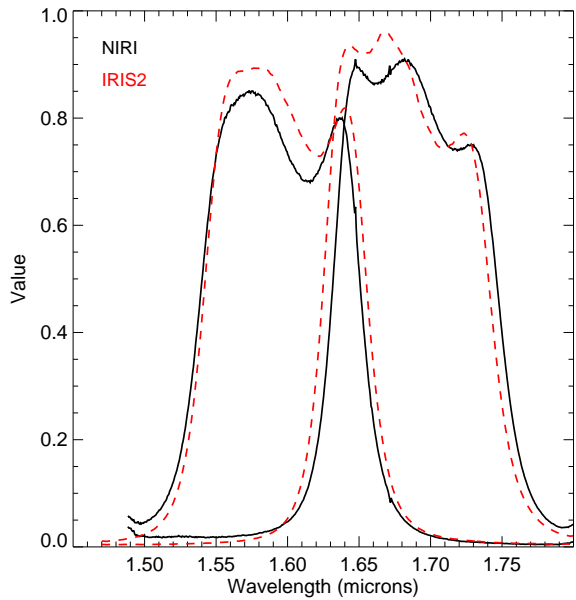}
   \caption{Profiles of the methane ON (long wavelength;
centered at 1.58$\mu$m) and OFF (short wavelength; centered
at 1.69$\mu$m) filters available on IRIS2 on the AAT (red
dashed lines) and NIRI on Gemini (black solid lines).
}
   \label{fig_dT:filter_curves}
\end{figure}

\citet{tinney05} presented a relation between the spectral type and the 
methane index given by the CH$_{4}$s$-$CH$_{4}$l colour for 15 T dwarfs 
spanning the full T class (see their Table 2). The methane index is 
unique for T dwarfs later than T4, making this technique robust for 
identifying and classifying T dwarfs. According to Table 2 in 
\citet{tinney05}, a clear difference in the methane index is observed
between T6.5 and T7, with a sharp decrease of 0.5 mag. We have applied 
this technique to UDXS2216 using the methane filters on Gemini.
Our methane index of $-$1.23 mag is consistent with the values
quoted by \citet{tinney05} for 
SDSSp J134646$-$003150 \citep[T6.5;][]{tsvetanov00} and 
2MASS J121711-031113 \citep[T7;][]{burgasser99}, yielding a spectral type 
of T6.5--T7 for UDXS2216\@. This result is in agreement with the 
latest spectral range inferred from the $J-K$ colour alone. 
Similarly, we estimate a spectral type of T6 from a methane index of 
$-$0.85$\pm$0.11 for UDXS2219, consistent with the indices derived by 
\citet{tinney05} for 2MASS J222828$-$431026 \citep[T6;][]{burgasser03c} 
and SDSSp J134646$-$003150 \citep[T6.5;][]{tsvetanov00}.
The difference between the colours obtained with the Gemini NIRI 
and Anglo-Australian Telescope IRIS2 instrument should be small 
($<$0.05 mag) because both sets of filters present similar profiles
(Fig.\ \ref{fig_dT:filter_curves}).

To summarise, we adopt spectral types of T7.0$\pm$1.0 and T6$\pm$1.0
for UDXS2216 and UDXS2219, respectively. Nonetheless, additional 
SDSS $z$-band, $Y$-band, and/or mid-infrared 4.5$\mu$m imaging would 
add further constraints on the spectral type and allow a direct 
comparison with a large number of late-T dwarfs from the LAS 
\citep{lodieu07b,pinfield08,burningham08a}.


%
%
%
%

%
%
\section{Discussion}
\label{dT:discussion}
%





%
\subsection{Photometric distances}
\label{dT:dist}

There have been eight T7 dwarfs published to date (see the L and T
dwarf archive)$^{1}$, including Gl229B classified as peculiar
\citep{nakajima95,burgasser06a} and only one with a measured parallax,
2MASS J072718$+$171001 \citep{burgasser02,vrba04}. Assuming an
absolute $J$ magnitudes (M$_{J}$) of 15.81 mag for the latter object and 
an dispersion of 0.5 mag on the spectral type vs M$_{J}$ relation, 
UDXS2216 would be located at 75$^{+20}_{-15}$ pc. This estimate is 
in agreement with the 
77 pc derived using the absolute magnitude of a T7 dwarf quoted 
by \citet{vrba04}. The spectral type vs M$_{J}$ relations 
given by \citet{knapp04}, \citet{liu06}, and \citet{looper08a} lead 
to comparable distance intervals of 73--118 pc, 61--113 pc and
52--96 pc, respectively (assuming that the 
object is single). Therefore, we adopt a mean photometric distance of 
81$_{-29}^{+36}$ pc, making UDXS2216 one of the coolest and furthest 
field T dwarfs along with NTTDF J1205$-$0744 
\citep[T7; $J \sim$ 20.15 mag and $J-K \sim -$0.15 mag;][]{cuby99}.

There are two T6 dwarfs (not resolved as binary systems) with parallaxes,
2MASSI J024313$-$245329 \citep[$J$ = 15.38; d = 10.68 pc][]{burgasser02} 
and SDSSp J162414$+$002915 \citep[$J$ = 15.494; d = 11 pc][]{strauss99}
along with the companion of SCR 1845$-$6357A \citep{biller06} located
at 3.85 pc \citep{henry06}. As nearby stars, those three objects have
well-determined distances and suggest distances in the 54--59 pc 
range for UDXS2219\@. The absolute magnitude relation given
by \citet{vrba04} suggests a distance of 60 pc. The spectral type 
vs M$_{J}$ relations given by \citet{knapp04}, \citet{liu06}, and 
\citet{looper08a} lead to comparable distance intervals of 54--86 pc, 
55--87 pc and 44--70 pc, respectively (assuming that the
object is single). Therefore, we adopt a mean value of 60$^{+27}_{-16}$ pc
based on these seven measurements. We note that low-resolution near-infrared 
spectroscopy of UDXS2219 should be possible with current instrumentation 
($J$ = 19.0 mag).


%
%
%
\begin{figure*}
  \includegraphics[width=\linewidth]{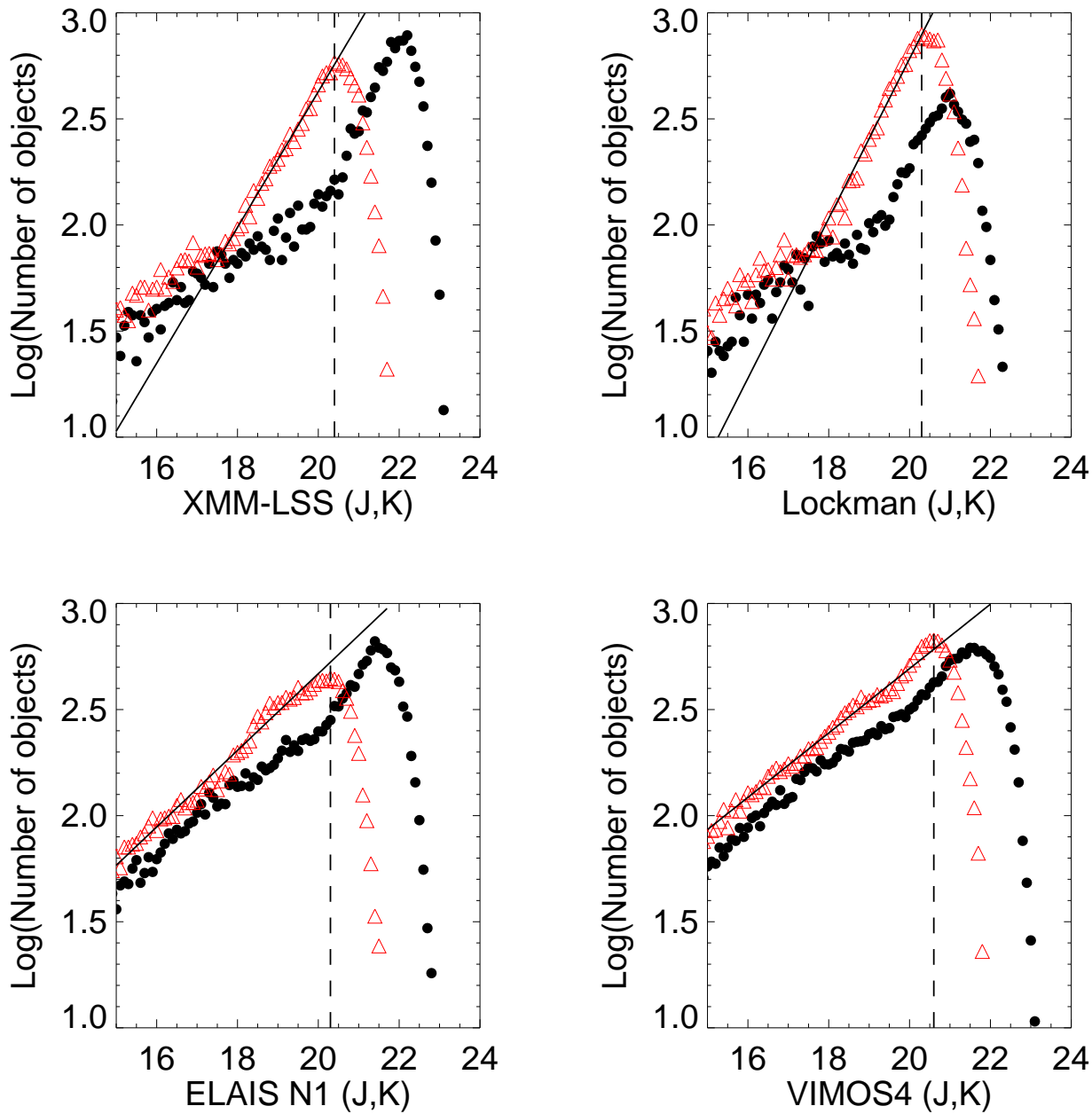}
   \caption{Histogram of the number of good quality point sources
per magnitude bin as a function of $J$ (filled circles) and $K$ 
(open triangles) in the four fields surveyed by the UKIDSS DXS: XMM-LSS
(top left), Lockman Hole (top right), ELAIS N1 (bottom left), and VIMOS 4
(bottom right). The histograms are scaled to a one-square-degree area.
Polynomial fits to the histograms are shown as solid
lines and 100\% completeness limits as dot-dashed lines (see numbers
in Table \ref{tab_dT:results_simulations}). 
   \label{fig_dT:fc_completeness}
}
\end{figure*}
%

%
%
%
\begin{table}
 \centering
 \caption[]{Simulated number of T dwarfs expected in each field observed
by the DXS for different values of the power law index $\alpha$ defined
as d$n$/d($\log$ M) = M$^{-\alpha}$ (in Salpeter units). The last line
give the number of expected T dwarfs in all four fields with a total
area of 6.16 deg$^{2}$ covered in $J$ and $K$ and an average depth
of $K \sim$ 20.4 mag.}
 \begin{tabular}{@{\hspace{0mm}}l c c c c c c@{\hspace{0mm}}}
 \hline
Field      & Area & $K$ depth & $-$1.5 & $-$1.0 & $-$0.5 &  0.0   \cr
\hline
Lockman    & 0.77 &   20.3    & 0.23   & 0.30   & 0.45   &  0.68  \cr
XMM-LSS    & 0.77 &   20.4    & 0.22   & 0.31   & 0.52   &  0.84  \cr
ELAIS\,N1  & 1.54 &   20.3    & 0.36   & 0.56   & 0.82   &  1.66  \cr
VIMOS\,4   & 3.08 &   20.6    & 1.16   & 1.80   & 2.58   &  4.18  \cr
\hline
All        & 6.16 &   20.5    & 1.97   & 2.97   & 4.37   &  7.36  \cr
\hline
 \label{tab_dT:results_simulations}
 \end{tabular}
\end{table}
\subsection{Expected number of T dwarfs in the DXS}
\label{dT:nb_DXS}

We have found two late-T dwarfs in six square 
degrees down to a completeness limit ranging from $K$ = 19.2 to 
$K$ = 21.1 mag (Sect.\ \ref{dT:selection_DXSDR2}). We have estimated
the depth of the DXS DR2 in each individual field for good quality point 
sources ({\tt{jClass}} and {\tt{kClass}} parameter between $-$2 
and $-$1; ellipticity less than 0.333; {\tt{jppErrBits}} and 
{\tt{kppErrBits}} parameter less than 256). This combination of
parameters provides an estimate of the point-source detection of
the data for single band detections only. The assessment of the
detection efficiency of our selection is, however, a strong function
of the types and colours of the sources being sought. Including
a colour cut to the above criteria would add different kind of biases.
For example, the colour selection made for our purpose 
($J-K \leq -$0.1 mag) will result in an extremely small statistics 
whereas the consideration of all sources will introduce a large sample 
of extremely red objects (mainly extragalactic) with a better detection 
in $K$ than in $J$. On the other hand, an intermediate colour criterion 
will return a complicated function of decreasing number of stellar 
sources and increasing number of extragalactic sources.

Figure \ref{fig_dT:fc_completeness} shows the histograms of the number
of point sources per magnitude bin as a function of the $J$ or 
$K$ magnitudes (scaled to a one-square-degree area) in the XMM-LSS, 
Lockman Hole, ELAISN N1, and VIMOS 4 fields. We infer a 100\% completeness 
limit of $K$ = 20.3--20.6 mag for point sources with a small variation 
across the four fields (Table \ref{tab_dT:results_simulations}). 
Many sources may however be extragalactic sources, especially at 
the faint end of the histograms. These limits are deduced from 
the points where the histograms deviate from the powerlaw fit to the counts
(Figure \ref{fig_dT:fc_completeness}).
As we are looking for blue objects, our search is 
currently limited by the depth in $K$. The total area surveyed in each 
field is 0.77, 0.77, 1.54, and 3.08 square degrees for the XMM-LSS, 
Lockman Hole, ELAISN N1, and VIMOS 4 field, respectively
(Table \ref{tab_dT:results_simulations}).

\citet{deacon06} provide simulations of the expected number of late-T 
dwarfs (defined as brown dwarfs cooler than 1300\,K) for various forms 
of the Initial Mass Function (IMF) and birthrates. We have adapted their 
simulations to the DXS and used a similar set of simulations to those 
used in \citet{pinfield08}. These are essentially those from 
\citet{deacon06} but with a few changes described now. The normalisation
factor i.e., the density of stars with masses $M$ between 0.09 and 
0.1 M$_{\odot}$, was set 0.0038 pc$^{-3}$ \citep{deacon08a}.
The uncertainty on that normalised factor is about 30\%. 
Also, the former simulations did not require a detection in $K$
but $J$ and $H$. However, for the DXS and UDS simulations, we have 
used only $J$ and $K$ (and not $Y$ and $H$) because those surveys are
only observing in those two passbands.
Additionally, the Galactic model in use is now the one published by 
\citet{deacon08a}. Finally, the number of simulations has been
increased by a factor of ten to get rid off the small number fluctuations.

Following these upgraded simulations,
the numbers of expected late-T dwarfs for different values of the 
$\alpha$ power law index (defined in the Salpeter scale as 
d$n$/d($\log$ M) = M$^{-\alpha}$) and scaled to the depth and area 
of each DXS field from DR2 are shown in 
Table \ref{tab_dT:results_simulations}. Thus, we should expect between
2.0 and 7.4 late-T dwarfs in 6.16 square degrees down to $K \sim$ 20.5 mag
for $\alpha$ indices of $-$1.0, $-$0.5, and 0 as well as a lognormal
form of the IMF (last line in Table \ref{tab_dT:results_simulations}). 
We should also mention that we looked only for candidates bluer than 
$J-K$ = $-$0.1 and some redder late-T dwarfs might have escaped our 
search criteria. The results agree with a declining mass function but 
statistics are too small to set any limits on underlying parameters of 
the IMF such as the $\alpha$ index.

The second object is fainter than the limits set for the simulations
and lies, within the photometric error bars, at the depth that the DXS
aimed to achieve for each tile. If we assume a depth of $K$ = 21.1 mag
as chosen for our selection search (Sect.\ \ref{dT:selection_DXSDR2}),
we would probe a volume 2.3 times larger (on average), implying that
we should find 4.6--17 late-T dwarfs in DXS DR2\@.
However, we are only 95\% complete in this extra magnitude bin 
($K$ = 20.5--21.1 mag) from the ratio of the extrapolated power 
law fit to the histogram shown in Fig.\ \ref{fig_dT:fc_completeness} 
to the observed number of objects. We found one T dwarf candidate
whereas the simulations predict 4.4--16.2\@.

\subsection{Search in the UDS DR2}
\label{dT:UDS_DR2}

The UDS field is centered on (2$^{\rm h}$18,$-$5$^{\circ}$10$'$) and
is located to the west of the XMM-LSS field in the DXS
\citep{lawrence07}.  It was chosen to overlap with the Subaru/XMM Deep
Survey field \citep{furusawa08} where multi-wavelength coverage is
available\footnote{More details at:
http://www.nottingham.ac.uk/astronomy/uds}. The UDS
covers one tile only and will be repeated over the 7 year UKIDSS plan
with a 3$\times$3 microstepping (pixel size of 0.1342 arcsec)
to achieve depths of $J$ = 25.0 mag , $H$ = 24.0 mag and $K$ = 24.0 mag
($5\sigma$ point-like sources) over a contiguous area of 0.77 deg$^2$.
Seeing constraints are 0.85 arcsec in $J$ with a sky brightness
greater than 16 mag/arcsec$^{2}$ and seeing at $K$ less than 0.75
arcsec with no sky brightness limit \citep{dye06}. The data processing
follows the standard WFCAM processing (Irwin et al., in prep.) up to
the creation of the intermediate deep stacks. The remaining steps to
create the final stacked tile is achieved with special processing
developed by the UDS team as described in \cite{foucaud07} and in more
detail in a forthcoming paper (Almaini et al., in prep.).  The data
releases used in this paper are DR2 \citep{warren07b} and DR3,
reaching depths of $J$ = 22.8 mag and
$K$ = 21.6 mag, and $J$ = 22.8 mag, $H$ = 22.1 mag and $K$ = 21.8 mag
respectively. The measured mean seeing is 0.90$''$ in $J$, 0.85$''$ 
in $H$ and 0.75$''$ in $K$. Internal astrometry is good with
a rms of 30 milli-arcsec. The stacking method used was a
weighted mean stacking method for DR2 and a 3$\sigma$ clipping weighted 
stacking method for DR3 which hampers the detection of
objects with proper motions larger than the resolution of the 
final stacked image.

We have applied the same constraints detailed in 
Sect.\ \ref{dT:selection_DXSDR2} to the UDS DR2: {\tt{jClass}}
and {\tt{kClass}} between $-$2 and $-$1, ellipticity less than 0.333,
and $J-K \leq -$0.1 mag. However, two differences should be emphasised:
first, the 5$\sigma$ limit of the UDS is deeper in $J$ and $K$ and we
have set it to $K$ = 21.5 mag; second, the UDS field-of-view is made solely
of one WFCAM tile i.e.\ $\sim$0.8 square degree located in the 
Subaru/XMM-Newton Deep Survey field. However, plotting the same histograms
as for the DXS suggests that the depth of the UDS DR2 for point sources is
$J \sim K \sim$ 20.0 mag (actually shallower than the DXS) because the
coverage do not overlap, implying that the UDS is not taking advantage of
the depth of the DXS\@. If we include galaxies, the depth
corresponds to the values quoted in WSA ($J$ = 21.6 mag and $K$ = 21.5 mag).
Our query returned eight candidates but all of them turned out to be 
cross-talks. Hence, no new T dwarf candidate was extracted. Moreover, 
we have extended this search to the latest data release (DR3) and didn't 
find any new candidate despite a survey about one magnitude deeper
in $J$ but not $K$.

Assuming the predictions by \citet{deacon06} for the UDS with
the upgrades detailed
in the previous section (Sect.\ \ref{dT:nb_DXS}), we should find at 
most three late-T dwarfs for the depth ($K \sim$ 20 mag) of the UDS DR2\@.
Those numbers are statistically in agreement with the non-detection of 
late-T dwarfs in the current data if we add the fact that we are likely
missing T dwarfs with large proper motions.
Upcoming UDS releases should allow us to extract (and possibly confirm)
some candidates, as the simulations predict around 40--200 late-T dwarfs,
depending on the shape of the mass function.

%
%
\section{Conclusions}
\label{dT:conclusions}

We have presented the discovery of two new late-T dwarfs at $\sim$60
and 80 pc (95\% confidence level) extracted photometrically from six 
square degrees surveyed by the UKIDSS DXS DR2\@. These new T dwarfs, 
UDXS2219 and UDXS2216, have $J-K$ colours and methane indices suggesting
spectral types of T6 and T7 (with an uncertainty of one subclass), 
respectively. These are the first confirmed late-T dwarfs found
in the DXS VIMOS 4 field. They are also among the coolest and most 
distant T dwarfs found to date. We have demonstrated the viability 
of a simple, but well chosen catalogue query for finding brown dwarfs.

These discoveries open new prospects and represent a first step to 
determine the scale height of 
T dwarfs. They are currently eight late-T dwarfs at distances larger than 
50 pc (this number is approximate since the distance estimates are subject
to an uncertainty on the spectral type vs M$_{J}$ relation for T dwarfs). 
Moreover, the availability of public deep optical surveys in those regions 
are extremely valuable to hunt for ultracool brown dwarfs. In this context, 
additional optical ($i$ or $z$) imaging is required to complement the 
SDSS Stripe 82 and cover the full DXS VIMOS 4 field.
The upcoming UKIDSS data releases will provide deeper data in the four 
fields surveyed as part of the DXS and UDS and will certainly trigger 
additional discoveries of even further T dwarfs and eventually Y dwarfs. 

%
%
\section*{Acknowledgments}

We thank the anonymous referee for his/her comments and suggestions
which improved significantly the original version of the paper.
N.R.D.\ is funded by NOVA and by NWO-VIDI grant 639.041.405 to Paul Groot.
We also would like to thank the Gemini HelpDesk, especially John
Holt, for the help with the NIRI data reduction.

This research has made use of the Simbad database and of NASA's
Astrophysics Data System Bibliographic Services (ADS).
Research has benefitted from the M, L, and T dwarf compendium housed
at DwarfArchives.org and maintained by Chris Gelino, Davy Kirkpatrick,
and Adam Burgasser.

The United Kingdom Infrared Telescope is operated by the Joint 
Astronomy Centre on behalf of the U.K. Science Technology and
Facility Council. 

Based on observations obtained at the Gemini Observatory (programs 
GN-2007A-Q-88 and GN-2008B-Q-90), which is operated by the 
Association of Universities for Research in Astronomy, 
Inc., under a cooperative agreement with the NSF on behalf of the 
Gemini partnership: the National Science Foundation (United States), 
the Particle Physics and Astronomy Research Council (United Kingdom), 
the National Research Council (Canada), CONICYT (Chile), the Australian 
Research Council (Australia), CNPq (Brazil) and SECYT (Argentina).
The SDSS is managed by the Astrophysical Research Consortium 
for the Participating Institutions. The Participating Institutions 
are the American Museum of Natural History, Astrophysical Institute 
Potsdam, University of Basel, University of Cambridge, Case Western 
Reserve University, University of Chicago, Drexel University, Fermilab, 
the Institute for Advanced Study, the Japan Participation Group, Johns 
Hopkins University, the Joint Institute for Nuclear Astrophysics, the 
Kavli Institute for Particle Astrophysics and Cosmology, the Korean 
Scientist Group, the Chinese Academy of Sciences (LAMOST), Los Alamos 
National Laboratory, the Max-Planck-Institute for Astronomy (MPIA), 
the Max-Planck-Institute for Astrophysics (MPA), New Mexico State 
University, Ohio State University, University of Pittsburgh, University 
of Portsmouth, Princeton University, the United States Naval Observatory, 
and the University of Washington.

This research has benefitted from observations obtained with 
MegaPrime/MegaCam, a joint project of CFHT and CEA/DAPNIA, at the 
Canada-France-Hawaii Telescope (CFHT) which is operated by the National 
Research Council (NRC) of Canada, the Institut National des Science 
de l'Univers of the Centre National de la Recherche Scientifique (CNRS) 
of France, and the University of Hawaii. This work is based in part on 
data products produced at TERAPIX and the Canadian Astronomy Data 
Centre as part of the Canada-France-Hawaii Telescope Legacy Survey, 
a collaborative project of NRC and CNRS.

This research used the facilities of the Canadian Astronomy Data 
Centre operated by the National Research Council of Canada with 
the support of the Canadian Space Agency.

%
%
\bibliographystyle{mn2e}
\bibliography{../../AA/mnemonic,../../AA/biblio_old}

\label{lastpage}

\end{document}